\documentstyle[seceq,preprint]{jpsj}
\input epsf.tex

\def\runtitle{
Enhanced Impurity Scattering due to Quantum Critical Fluctuations}
\def\runauthor{
Kazumasa {\sc Miyake} and Osamu {\sc Narikiyo}
}

\title{
Enhanced Impurity Scattering due to Quantum Critical Fluctuations: 
Perturbational Approach 
}

\author{
Kazumasa {\sc Miyake} and Osamu {\sc Narikiyo}$^{a)}$
}

\inst
{
Department of Physical Science, Graduate School of Engineering Science, 
Osaka University, Toyonaka, Osaka 560-8531; 

$^{a)}$Department of Physics, Graduate School of Science, 
Kyushu University, Fukuoka 812-8581
}
\recdate
{
May 26, 2000
}

\abst
{
It is shown on the basis of the lowest order perturbation expansion 
with respect to critical fluctuations that the critical fluctuations 
give rise to an 
enhancement of the potential scattering of non-magnetic impurities.  
This qualitatively accounts for the 
enhancement of the resistivity due to impurities which has been 
observed in variety of systems near the quantum critical point, while 
the higher order processes happen to give much larger enhancement 
as seen from the Ward identity arguments.  
The cases with dynamical critical exponent $z$=2 and $z$=3 are discussed 
explicitly.  
}

\kword
{enhanced impurity scattering, quantum critical phenomena, 
enhanced residual resistivity
}

\begin{document}
\sloppy
\maketitle
\section{Introduction}
Recently the so-called non-Fermi liquid behaviors near the 
quantum critical point (QCP) have attracted much attentions 
in heavy electron systems in which the tuning of the quantum 
parameter is relatively easy
\cite{lohneysen,takabatake,miyako,kambe,lonzarich1,
lonzarich2,tckobayashi,jaccard2,steglich,maezawa}, 
while such phenomena have also been studied in $d$-band metals 
exhibiting ferromagnetic QCP
\cite{smith, pfleid,thessieu}.  
Indeed, pressures of the order of 
several GPa can convert the magnetic ground state into the 
paramagnetic one, or {\it vice versa}, through the magnetic QCP
\cite{takabatake,lonzarich1,lonzarich2,tckobayashi,jaccard2,steglich,
trovarelli,knebel}.  
The anomalous behaviors of physical quantities around metallic and magnetic 
QCP have been successfully analyzed by the so-called SCR theory 
(and its extentions) \cite{kambe,moriya,moriya2}, except for 
some cases\cite{lohneysen,trovarelli}, 
while the basis of SCR theory 
has also been established on the basis of perturbational renormalization 
group method at least up to the intermediate coupling regime
\cite{hertz,millis}.  The reason why such an approach is successful even in 
the strongly correlated metals should be attributed to the validity of the 
Fermi liquid theory\cite{landau} for the description of 
the normal state on which the magnetic phase transition can be discussed.  

It has been well recognized that the impurity scattering gives rise to 
a drastic effect on the critical behaviors near the QCP\cite{moriya}.  
For example, the dynamical exponent $z$ is altered considerably 
owing to the impurity scattering: $z$=3 in pure system for the 
ferromagnetic QCP is changed to $z$=4 in the system with non-magnetic 
impurities\cite{impurity}.  The effect of impurity scattering on 
the critical exponent of the temperature dependence of the resistivity 
has recently been discussed and attracted much interests~\cite{rosch}.  
However, the effect of critical fluctuations on the impurity 
potential has scarcely been discussed so far, while a general 
argument on the Ward identity was given in relation to the insulating 
behaviors in non-Fermi liquids subject to the impurity scattering
\cite{kotliar,varma}.  On the other hand, the enhancement of 
the residual resistivity $\rho_{0}$ at around QCP has been reported 
in MnSi under the pressue of $P\simeq15$GPa\cite{thessieu} where 
the ferromagnetic state disappears, and in CeCu$_{2}$Ge$_{2}$ under 
the pressue of $P\simeq17$GPa where the rapid valence change of Ce 
ion may occur\cite{jaccard2,miyake}.  
Similar behavior has recently been observed in 
UGe$_2$ under pressures\cite{oomi}.  
It has also been reported that the resistivity of 
CeNi$_{2}$Ge$_{2}$ under $P\simeq1.6$GPa grows up when the temperature 
is decreased below $T=2$K~\cite{maezawa}.  

A purpose of this paper is to discuss the effect of quantum critial 
fluctuations on the potential of non-magnetic impurity.  It is shown 
on the basis of the perturbational treatment that the fluctuations 
associated with QCP can give rise to an enhancement of impurity 
potential which is , in the 
limit of zero-momentum transfer, in proportional to the mass enhancement 
factor 1/$z_{\rm cr}$, 
due to the critical fluctuations.  As a result, 
the residual resistivity $\rho_{0}$ is shown to exhibit an 
enhancement as a function of the parameter measuring quantum criticality.  
This consequence is quite general while 
the mass enhancement factor itself depends on the dynamical 
exponent, and can be compared to the experimental observations 
in a couple of systems exhibiting QCP.   

\section{Renormalization of impurity potential by critical fluctuations \\
- A perturbational approach -}
\begin{figure}[htbp]
\begin{center}
\epsfxsize=3.5cm \epsfbox{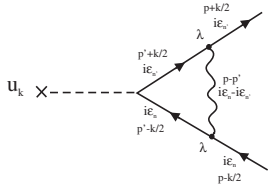}
\end{center}
\caption{Feynman diagram for the vertex correction of impurity 
potential $u_{q}$ by exchanging one fluctuation mode. 
}
\label{fig:1}
\end{figure}

In this section, we discuss the renormalization of impurity potential due to 
exchanging critical fluctuations by perturbation method.  
The lowest order correction of impurity potential is given by the 
Feynman diagram shown in Fig.\ \ref{fig:1}.  Its analytic expression for 
the vertex correction factor, 
corresponding to the scattering of quasiparticles from 
${\vec p}-{\vec k}/2$ to ${\vec p}+{\vec k}/2$, is 
given as 
\begin{eqnarray}
\label{eq:1}
\Delta\Gamma_{{\vec k},{\vec p}}({\rm i}\epsilon_{n};\eta)
&=&\lambda^{2}T\sum_{n^{\prime}}\sum_{{\vec p}^{\prime}}
\chi({\vec p}-{\vec p}{\rm \ }^{\prime},{\rm i}\epsilon_{n}-{\rm i}\epsilon_{n^{\prime}})
\nonumber
\\
&&\quad \times G({\vec p}{\rm \ }^{\prime}-{{\vec k}\over 2},
{\rm i}\epsilon_{n^{\prime}})
G({\vec p}{\rm \ }^{\prime}+{{\vec k}\over 2},{\rm i}\epsilon_{n^{\prime}}),
\end{eqnarray}
where $\lambda$ is the coupling constant between the critical fluctuation 
modes and the quasiparticles.  The Green function $G$ of 
quasiparticles is expressed as
\begin{equation}
\label{eq:2}
G({\vec p},{\rm i}\epsilon_{n})=\int_{-\infty}^{\infty}{\rm d}x
{A({\vec p},x)\over x-{\rm i}\epsilon_{n}},
\end{equation}
and the propagator of fluctuaion modes is expressed as 
\begin{equation}
\label{eq:3}
\chi({\vec q},{\rm i}\omega_{n})=\int_{-\infty}^{\infty}{\rm d}y
{B({\vec q},y)\over y-{\rm i}\omega_{n}}. 
\end{equation}
Here, we assume that the {\it bare} quasiparticles are well defined 
so that their specral weight is approximated as 
\begin{equation}
\label{eq:4}
A({\vec p},x)\simeq {\bar z}\delta(x-\xi_{{\vec p}}),
\end{equation}
where ${\bar z}$ is the renormalization amplitude due to the effects 
other than the critical fluctuations; namely, it includes the effect of 
local correlation leading to heavy electron state.  
The spectral weight of fluctuation mode is given, by definition, as 
\begin{equation}
\label{eq:5}
B({\vec q},y)\simeq{1\over \pi}{\rm Im}\chi({\vec q},y+{\rm i}\delta),
\end{equation}
where the propagator of fluctuation modes is assumed to be 
parametrized as follows:
\begin{equation}
\label{eq:6}
\chi({\vec q},\omega)\simeq
{\chi_{Q}^{(0)}\over
\eta+A({\vec q}-{\vec Q})^{2}-iC_{q}\omega},
\end{equation}
where $\chi_{Q}^{(0)}$ is of the order of ${\bar N}_{\rm F}$, the 
renormalized density of states (DOS) of {\it bare} 
quasiparticles at the Fermi level.  
The coefficient $C_{q}$ in (\ref{eq:6}) depends on $q$ in general 
and its $q$-dependence 
is dependent on the dynamical structure of QCP.  
In the case of the conventional antiferromagnetic (AF) fluctuations 
$C_{q}$ is eassentially independent of $q$ leading to the dynamical 
exponent $z$=2, while in the case of 
ferromagnetic fluctuations $C_{q}=C/q$, $C$ being a 
constant, leading to $z$=3\cite{moriya}.  
In the case of AF fluctuations where the even number of magnetic 
ions are contained in the unit cell and they are equivalent locally
\cite{hatatani}, and 
of uniform fluctuations which is not accompanied by conserved 
quantity\cite{varma2} such as valence fluctuations\cite{miyake}, 
$C_{q}=C/{\rm max}\{q, \ell^{-1}\}$ in the limit $T\to 0$, 
$\ell$ being the mean free path of the impurity scattering.  

Substituting (\ref{eq:2}) and (\ref{eq:3}) into (\ref{eq:1}) and 
performing the summation with respect to $n^{\prime}$, 
the vertex correction factor for scattering potential is reduced to 
\begin{eqnarray}
\label{eq:7}
\Delta\Gamma_{{\vec k},{\vec p}}(\epsilon+{\rm i}\delta;\eta)
&=&{{\tilde \lambda}^{2}\over 2}\sum_{{\vec q}}
\int_{-\infty}^{\infty} {\rm d}yB({\vec q},y)
{1\over \xi_{{\vec p}-{\vec k}/2-{\vec q}}-
\xi_{{\vec p}+{\vec k}/2-{\vec q}}}\nonumber\\
&&\times\Biggl[{\displaystyle\strut \coth{y\over 2T}+
\tanh{\xi_{{\vec p}+{\vec k}/2-{\vec q}}\over 2T}\over
y+\xi_{{\vec p}+{\vec k}/2-{\vec q}}-\epsilon-{\rm i}\delta}
-{\displaystyle\strut \coth{y\over 2T}+
\tanh{\xi_{{\vec p}-{\vec k}/2-{\vec q}}\over 2T}\over
y+\xi_{{\vec p}-{\vec k}/2-{\vec q}}-\epsilon-{\rm i}\delta}\Biggr],
\end{eqnarray}
where the analytic continuation 
${\rm i}\epsilon_{n}\to\epsilon+{\rm i}\delta$ 
has been performed, and 
${\tilde \lambda}\equiv {\bar z}\lambda$ is the effective 
coupling constant.  In the limit of $T\to 0$, 
using the spectral function $B$, (\ref{eq:5}), 
with $\chi$ given by (\ref{eq:6}), the integration with respect to 
$y$ in (\ref{eq:7}) can be easily performed obtaining 
\begin{equation}
\label{eq:8}
\Delta\Gamma_{{\vec k},{\vec p}}(\epsilon;\eta)
={{\tilde \lambda}^{2}\over 2\pi}\sum_{{\vec q}}
\chi({\vec q},0)
{F({\tilde E}_{+},\xi_{+})-
F({\tilde E}_{-},\xi_{-})\over \xi_{-}-\xi_{+}}
\end{equation}
where $\xi_{\pm}\equiv \xi_{{\vec p}\pm{\vec k}/2-{\vec q}}$ , and 
the fuctions F's are defined by 
\begin{equation}
\label{eq:9}
F({\tilde E}_{\pm}(\epsilon),\xi_{\pm})
\equiv {-2{\tilde E}_{\pm}\ln |{\tilde E}_{\pm}|+
\pi{\rm sign}(\xi_{\pm})
\over{\tilde E}_{\pm}^{2}+1},
\end{equation}
where 
\begin{equation}
\label{eq:10}
{\tilde E}_{\pm}(\epsilon)\equiv 
{C_{q}(\epsilon-\xi_{\pm})\over
\eta+Aq^{2}}.
\end{equation}
  
\begin{figure}[htbp]
\begin{center}
\epsfxsize=3.5cm \epsfbox{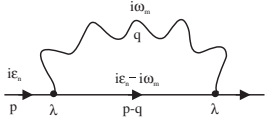}
\end{center}
\caption{Feynman diagram for the selfenergy by exchanging one 
fluctuation mode. 
}
\label{fig:2}
\end{figure}

Before examining the $\eta$ dependence of (\ref{eq:8}), 
let us investigate the lowest order correction for the 
selfenergy of quasiparticles due to the critical 
fluctuations given by (\ref{eq:6}).  Such a selfenergy 
is given by the Feynman diagram shown in Fig.\ \ref{fig:2} 
which corresponds to Fig.\ \ref{fig:1} for the lowest order 
correction to the impurity potential.  Analytic expression 
of this selfenergy is given by 
\begin{equation}
\label{eq:11}
\Sigma({\vec p},{\rm i}\epsilon_{n})=
\lambda^{2}T\sum_{m}\sum_{{\vec q}}
\chi({\vec q},{\rm i}\omega_{m})
G({\vec p}-{\vec q},{\rm i}\epsilon_{n}-{\rm i}\omega_{m}).
\end{equation}
Performing the summation with respect to $m$ and analytic continuation 
${\rm i}\epsilon_{n}\to \epsilon+{\rm i}\delta$, we obtain
\begin{eqnarray}
\label{eq:12}
{\bar z}\Sigma({\vec p},\epsilon+{\rm i}\delta)
=-{{\tilde \lambda}^{2}\over 2}&&\sum_{{\vec q}}
\int_{-\infty}^{\infty}{\rm d}yB({\vec q},y)\nonumber\\
&&\times
{\displaystyle\strut \coth{y\over 2T}+
\tanh{\xi_{{\vec p}-{\vec q}}\over 2T}\over 
y+\xi_{{\vec p}-{\vec q}}-\epsilon-{\rm i}\delta}.
\end{eqnarray}
In the limit of $T\to 0$, using (\ref{eq:5}) and (\ref{eq:6}), 
$y$-integration is easily performed 
obtaining 
\begin{equation}
\label{eq:13}
{\bar z}\Sigma({\vec p},\epsilon+{\rm i}\delta)
=-{{\tilde \lambda}^{2}\over 2\pi}\sum_{{\vec q}}
\chi({\vec q},0)F_{0}({\tilde E}_{0},\xi_{{\vec p}-{\vec q}}),
\end{equation}
where the function $F_{0}$ is defined as 
\begin{equation}
\label{eq:14}
F_{0}({\tilde E}_{0}(\epsilon),\xi_{{\vec p}-{\vec q}})\equiv
{-2{\tilde E}_{0}\ln |{\tilde E}_{0}|
+\pi{\rm sign}(\xi_{{\vec p}-{\vec q}})\over
{\tilde E}_{0}^{2}+1},
\end{equation}
where 
\begin{equation}
\label{eq:15}
{\tilde E}_{0}(\epsilon)\equiv 
{C_{q}(\epsilon-\xi_{{\vec p}-{\vec q}})
\over
\eta+Aq^{2}}.
\end{equation}  
It is noted that the following relation holds: 
\begin{equation}
\lim_{k\to 0}F({\tilde E}_{\pm}(\epsilon),\xi_{\pm})
=F_{0}({\tilde E}_{\pm}(\epsilon),\xi_{{\vec p}-{\vec q}}).
\label{eq:15s}
\end{equation}  
From Eqs. (\ref{eq:13})-(\ref{eq:15}), we obtain the relation
\begin{eqnarray}
\label{eq:16}
-{\bar z}{\partial\Sigma({\vec p},\epsilon)\over
\partial\epsilon}
&=&
{{\tilde \lambda}^{2}\over 2\pi}\sum_{{\vec q}}
\chi({\vec q},0)\nonumber\\
&&\times{C_{q}\over \eta+Aq^{2}}
{\partial F_{0}({\tilde E}_{0},\xi_{{\vec p}-{\vec q}})\over
\partial{\tilde E}_{0}}.
\end{eqnarray}

In the limit of forward scattering, i.e., $k\to 0$, of the quasiparticles 
near the Fermi surface, the vertex correction factor (\ref{eq:8}) can be 
estimated with using (\ref{eq:9}) and (\ref{eq:10}) as follows:
\begin{eqnarray}
\label{eq:17}
\lim_{k\to 0}
\Delta\Gamma_{{\vec k},{\vec p}}(\epsilon;\eta)
&=&{{\tilde \lambda}^{2}\over 2\pi}\sum_{{\vec q}}
\chi({\vec q},0){C_{q}\over \eta+Aq^{2}}
\lim_{k\to 0}
{\partial F({\tilde E}_{\pm},\xi_{\pm})\over
\partial{\tilde E}_{\pm}}\nonumber\\
&&+{{\tilde \lambda}^{2}\over 2\pi}\sum_{{\vec q}}
\chi({\vec q},0)\lim_{k\to 0}
{\pi\over\xi_{-}-\xi_{+}}
\Biggl[{{\rm sign}(\xi_{+})-{\rm sign}(\xi_{-})\over
(\lim_{k\to 0}{\tilde E}_{\pm})^{2}+1}\Biggr].
\end{eqnarray}
The first term of (\ref{eq:17}) is equal to 
$-{\bar z}\partial\Sigma({\vec p},\epsilon)/
\partial\epsilon$ given by (\ref{eq:16}) owing to the relation 
(\ref{eq:15s}).  The second term of (\ref{eq:17}) 
vanishes, because the phase space of ${\vec q}$ satisfying the condition 
${\rm sign}(\xi_{+}){\rm sign}(\xi_{-})<0$ is restricted in a very narrow 
region as can be seen by geometrical consideration in the wave vector 
space.  Indeed, its component $\Delta q_{\parallel}$ parallel to 
${\vec p}\approx{\vec p}_{\rm F}$ is restricted in the region 
$|\Delta q_{\parallel}|<(k/p_{\rm F})^{2}$ because the angle between 
${\vec p}-{\vec k}/2$ and ${\vec p}+{\vec k}/2$ is proportional to 
$(k/p_{\rm F})^{2}$ , so that the second term of 
(\ref{eq:17}) vanishes as $\propto k$.  Therefore, we obtain 
\begin{equation}
\label{eq:18}
\lim_{k\to 0}
\Delta\Gamma_{{\vec k},{\vec p}}(\epsilon;\eta)
\simeq
-{\bar z}{\partial\Sigma({\vec p},\epsilon)\over
\partial\epsilon},
\end{equation}
which implies that the renormalized impurity potential 
${\tilde u}({\vec k})$ is 
given, in the zero momentum transfer limit, as 
\begin{equation}
\label{eq:19}
{\tilde u}({\vec k}\to 0;{\vec p})=
\biggl[1-{\bar z}{\partial\Sigma({\vec p},\epsilon)\over
\partial\epsilon}\biggl]u({\vec k}\to 0),
\end{equation}
where $u({\vec k})$ is the bare impurity potential.  Namely, ${\tilde u}$ 
is enhanced by the mass enhancement factor 
$1/z_{\rm cr}({\vec p},\epsilon)
\equiv[1-{\bar z}\partial\Sigma({\vec p},\epsilon)/
\partial\epsilon)]$ which expresses the excess enhancement 
due to the critical fluctuations beyond the local correlations leading 
to the heavy electrons.  
This is consistent with the exact result obtained on the Ward identity 
argument by Betbeder-Matibet and Nozi\`eres\cite{matibet}, 
who showed on the Fermi liquid formalism that the renormalized impurity 
potential is given as 
\begin{equation}
\label{eq:20}
{\tilde u}({\vec k}\to 0)={1\over z(1+F_{0}^{\rm s})}
u({\vec k}\to 0),
\end{equation}
where $z$ is the renormalization amplitude including {\it all} the manybody 
effects and $F_{0}^{\rm s}$ the Landau parameter.  
Fermi liquid correction corresponds to that of higher order perturbation 
with respect to critical fluctuations which is beyond treatment 
in this paper.  It is noted that renormalized impurity potenital depends 
on the momentum ${\vec p}$ of incomming quasiparticles in general, 
especially near the AF-QCP.  
Of the effects making $z$ decrease, that arising from the 
local spin correlations should be cancelled by the factor 
$(1+F_{0}^{\rm s})$ in the heavy electrons as discussed by, 
e.g., in Ref. \cite{varma3}.  
However, those beyond it, such as valence or magnetic 
fluctuations associated with quantum criticality, can give rise to excess 
reduction of $z$.   

The renormalization effect of impurity potential given here is in 
a close relation to the conventional renormalization effect in the Fermi liquid 
theory for the conserved quantities which are expressed in terms of 
quasiparticles by the same formula as the non-interacting particles, 
i.e., the weight $z$ of quasiparticles in the one-particle spectral 
weight is cancelled by the vertex correction $z^{-1}$ due to the incoherent 
processes\cite{AGD,Leggett}.  The physical reason of the enhancement of 
the impurity potential may be understood as follows: According to 
the expression (\ref{eq:7}), the vertex correction includes the factor 
$B({\vec q},y)/(y+\xi_{{\vec p}\pm{\vec k}-{\vec q}}-\epsilon)$, where it is 
to be remembered that $B({\vec q},y)$ is the spectral weight of 
spin fluctuations with the wavevector ${\vec q}$ and the energy $y$.  
So, the enhancement may be related to that of the intermediate 
states of critical magnetic fluctuations associated with QCP.  

For a explicit calculation of the renormaliztion amplitude $z_{\rm cr}$, 
it is convenient to rewrite (\ref{eq:16}) directly from ({\ref{eq:12}) 
in the form\cite{maebashi}
\begin{equation}
\label{eq:21}
-{\bar z}{\partial\Sigma({\vec p},\epsilon)\over
\partial\epsilon}
={{\tilde \lambda}^{2}\over 4\pi^{3}}\int_{{\rm FS}}
{{\rm d}^{2}p{\rm \ }^{\prime}
\over |{\vec v}_{{\vec p}{\rm \ }^{\prime}}|}
{\rm Re}\chi({\vec p}-{\vec p}{\rm \ }^{\prime},\epsilon),
\end{equation}
where ${\vec v}_{{\vec p}'}$ 's are the velocity of {\it bare} 
quasiparticles, and FS indicates that the integration with respect to 
${\vec p}{\rm \ }^{\prime}$ is taken on the surface with 
$\xi_{{\vec p}^{\prime}}\approx\epsilon$.  
Using the explicit form (\ref{eq:6}) for the propagator of fluctuations 
with ${\vec Q}=0$ and the dynamical critical exponent $z$=3, 
the left hand side of (\ref{eq:21}) is 
calculated resulting in 
\begin{equation}
\label{eq:22}
-{\bar z}{\partial\Sigma({\vec p},\epsilon)\over
\partial\epsilon}
={{\tilde \lambda}^{2}\chi_{Q}^{(0)}\over 8\pi^{2}A\langle 
v_{\rm F}\rangle}
\cases{
\displaystyle{\strut 
\ln{Aq_{\rm c}^{2}+\eta \over\eta}
},  &$(\epsilon=0)$;\cr
\quad &\ \cr
\displaystyle{\strut 
{2\over 3}\ln{Aq_{\rm c}^{3}+C|\epsilon| \over C|\epsilon|}
},  &$(\eta=0)$,\cr}
\end{equation}
where $\langle v_{{\rm F}}\rangle$ is the averaged velocity of 
quasiparticles on the Fermi surface, and $q_{c}$ is a cut-off 
wavenumber of the order of 
inverse of the lattice constant.

Similarly, for a class of fluctuations with ${\vec Q}\not=0$ and 
the dynamical critical exponent $z$=2, 
$-{\partial\Sigma({\vec p},\epsilon)/\partial\epsilon}$ given by 
(\ref{eq:21}) depends crucially on the position of ${\vec p}$.  
Namely, for the momentum ${\vec p}_{\rm h}$ on the so-called 
``hot line", where 
$\xi_{{\vec p}_{\rm h}+{\vec Q}}=\xi_{{\vec p}_{\rm h}}=0$ on the 
Fermi surface, essentially the same expression as (\ref{eq:22}) is 
obtained.  Its $p$-dependence around the ``hot line" is parameterized 
by replacing $\eta$ by $\eta+Aq_{\rm m}^{2}$ in (\ref{eq:22}), where 
$q_{\rm m}$ is a measure of distance from the ``hot line" on the 
Fermi surface.   After averaging over $q_{\rm m}$ on the Fermi surface, 
one obtains 
\begin{eqnarray}
\label{eq:23}
-\Biggl\langle {\bar z}&&{\partial\Sigma({\vec p},\epsilon)\over
\partial\epsilon}\Biggr\rangle_{{\rm FS}}
\nonumber\\
&&={{\tilde \lambda}^{2}\chi_{Q}^{(0)}\over 8\pi^{2}A
\langle v_{\rm F}\rangle}
\cases{
\displaystyle{\strut 
b_{1}-b_{2}\sqrt{{\eta\over Aq_{\rm c}^{2}+\eta}}
},  &$(\epsilon=0)$;\cr
\quad &\ \cr
\displaystyle{\strut 
b_{1}-b_{2}\sqrt{{C|\epsilon|\over 2Aq_{\rm c}^{2}+C|\epsilon|}}
},  &$(\eta=0)$,\cr
}
\end{eqnarray}
where $b_{1}$ and $b_{2}$ are positive constant of 
${\cal O}(1)$ depending on the details of ${\vec Q}$ and 
shape of the Fermi surface.  
It is remarked that the results (\ref{eq:22}) and (\ref{eq:23}) 
are consistent with those for the specific heat anomaly of 
SCR theroy\cite{moriya, moriya2}.  
In any case, the impurity 
potenial for the forward scattering is enhanced in proportional 
to $1/z_{{\rm cr}}({\vec p},\epsilon)$ as given by (\ref{eq:19}).  

\section{Effect of critical fluctuations on residual resistivity}
In order to see how this enhancement of impurity potential 
affects the behaviors of the resistivity, we need to know the 
$k$-dependence of ${\tilde u}(k)$ for the scattering from 
${\vec p}-{\vec k}/2$ to ${\vec p}+{\vec k}/2$ near the Fermi 
surface.  At first sight, the equality 
(\ref{eq:18}) holds also for general values of momentum transfer ${\vec k}$ 
because the factor 
$[{F({\tilde E}_{+},\xi_{+})-F({\tilde E}_{-},\xi_{-})]
/(\xi_{-}-\xi_{+}})$ in (\ref{eq:8}) could be approximated 
by $-(\partial F/\partial {\tilde E})_{\xi=\xi_{\pm}}$  
for ${\vec q}\sim{\vec Q}$ where $\chi({\vec q},0)$ is diverging 
as $1/\eta$ near the QCP.  However, this is not the case as shown by 
explicit calculation of (\ref{eq:8}) without using such an 
approximation.
\footnote{
In this sense, a part of the results of Ref. \cite{miyake} should be 
revised as below.}  
Indeed, the $k$-dependence of ${\tilde u}(k)$ is estimeted as follows.  
The expansion of $[{F({\tilde E}_{+},\xi_{+})-F({\tilde E}_{-},\xi_{-})]
/(\xi_{-}-\xi_{+}})$ with respect to $k$ in (\ref{eq:8}) is allowed 
so long as 
$k\ll |{\vec p}-{\vec q}|\sim p_{\rm F}$, so that the relation similar 
to (\ref{eq:17}) holds.  However, the factor 
$\lim_{{\vec k}\to 0}\partial F({\tilde E}_{\pm},\xi_{\pm})/
\partial {\tilde E}_{\pm}$ in (\ref{eq:17}) should be replaced by 
$\partial F({\tilde E}_{\pm},\xi_{{\vec p}-{\vec q}})/
\partial {\tilde E}_{\pm}|_{{\tilde E}_{\pm}={\tilde E}_{0}(\epsilon-sk^{2})}
$ = $\partial F_{0}({\tilde E}_{0},\xi_{{\vec p}-{\vec q}})/
\partial {\tilde E}_{0}|_{{\tilde E}_{0}={\tilde E}_{0}(\epsilon-sk^{2})}$, 
where $s$ is a coefficient of ${\cal O}(1/{\bar m})$ with ${\bar m}$ being 
the effective mass of {\it bare} quasiparticles.  
It is because the energy of {\it bare} quasiparticle with momentum 
${\vec p}$ is given by 
$\xi_{p}\approx \xi_{{\vec p}\pm{\vec k}/2}-sk^{2}$ 
and different from those of incoming and outgoing particles, i.e., 
$\xi_{{\vec p}+{\vec k}/2}$=$\xi_{{\vec p}-{\vec k}/2}$, where we are 
considering the elastic scattering.  Therefore, for $k\ll p_{\rm F}$, 
we obtain instead of (\ref{eq:18}) the following relation 
\begin{equation}
\label{eq:24}
\Delta\Gamma_{{\vec k},{\vec p}}(\epsilon;\eta)
\simeq
-{\bar z}{\partial\Sigma({\vec p},\varepsilon)\over
\partial\varepsilon}\Biggr|_{\varepsilon=\epsilon-sk^{2}}.
\end{equation}
Namely, the impurity potential ${\tilde u}(k)$, for $k\ll p_{\rm F}$, 
giving the scattering on the Fermi surface (i.e., $\epsilon$=0), 
is renormalized as 
\begin{equation}
\label{eq:25}
{\tilde u}({\vec k};{\vec p})=
\biggl[1-{\bar z}{\partial\Sigma({\vec p},\varepsilon)\over
\partial\varepsilon}\biggl]_{\varepsilon=-sk^{2}}u({\vec k};{\vec p}),
\end{equation}
where $\partial\Sigma({\vec p},\varepsilon)/\partial\varepsilon$ is given by 
(\ref{eq:22}) in the case of QCP with $z$=3.  In the case of QCP with 
$z$=2, $\partial\Sigma({\vec p},\varepsilon)/\partial\varepsilon$ has large 
dependence on ${\vec p}$ as discussed above.  

If the bare impurity potential causes essentially the Born 
scattering, the residual resistivity $\rho_{0}$ is enhanced by 
the critical fluctuations.  Namely, $\rho_{0}$ is given as 
\begin{eqnarray}
\rho_{0}& &\propto
\Bigl\langle[1+
\Delta\Gamma_{{\vec k},{\vec p}}(0;\eta)
\bigr]^{2}\nonumber\\
& &\times 2\pi {\bar N}_{\rm F}{\bar z} 
c_{\rm imp}|u({\vec k};{\vec p})|^{2}(1-\cos \theta)\Bigr\rangle_{\rm FS},
\label{eq:26}
\end{eqnarray}
where $c_{\rm imp}$ is a concentration of impurity, 
$\theta$ is an angle between 
${\vec p}\pm{\vec k}/2$, and the on-shell 
condition $\epsilon$=$\xi_{{\vec p}\pm{\vec k}/2}$=0 has been used.  
Here it is noted that explicit dependence of 
renormalization amplitude $z_{\rm cr}$ does not appear due to 
cancellation between that for DOS and that for the damping rate of 
quasiparticles.  We perform the calculation of 
(\ref{eq:26}) for the spherical Fermi surface with 
using a unit of wavenumber such that $2p_{\rm F}\sqrt{s}$=1, and 
assuming that $Aq_{\rm c}^{2}$=1.  In the case of $z$=2, the average over 
$q_{\rm m}$, the measure of distance from the ``hot line", is approximated by 
$\int_{0}^{q_{\rm c}}{\rm d}q_{\rm m}(\cdots)/
\int_{0}^{q_{\rm c}}{\rm d}q_{\rm m}$.  
The results of numerical calculations of $\rho_{0}$ as a function of 
$\eta$ are shown in 
Fig.\ \ref{fig:3} in an arbitrary unit by normalizing at $\eta$=0.  
Although the logarithmic divergence 
in $\Delta\Gamma$ of (\ref{eq:22}) is smeared out due to the geometrical 
factor $(1-\cos\theta)$ in the case of $z$=3, a rather sharp cusp structure 
still remains in $\rho_{0}$ as a function of $\eta$, like 
$\rho_{0}\propto 1-8\eta$ for $\eta\sim 0$.   
Here it is noted that $k\approx 2p_{\rm F}\sin(\theta/2)$.  
In case of $z$=2, the result is much more smeared by 
the extra average process over $q_{\rm m}$.  

\begin{figure}[htbp]
\begin{center}
\epsfxsize=5.5cm \epsfbox{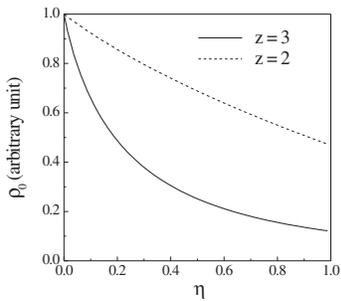}
\end{center}
\caption{Residual resistivity $\rho_{0}$ due to non-magnetic impurity 
as a function of inverse susceptibility $\eta$.  
}
\label{fig:3}
\end{figure}

\section{Discussions}
The result obtained in \S3 explains qualitatively the anomaly 
of $\rho_{0}$ observed near the ferromagnetic QCP 
\cite{thessieu,oomi},  because the main anomaly arises from the 
factor $1/z$ in (\ref{eq:20}).  
In the case of AF-QCP,  the less pronounced anomalies are expected to be 
observed.  This is also consistent with the experimental fact that 
no strong anomaly of $\rho_{0}$ is observed around AF-QCP with $z$=2.  

However, it needs careful consideration in the case of AF-QCP in compounds 
which have even number of magnetically equivalent ions in the primitive 
cell such as CeCu$_{6-x}$Au$_{x}$, containing magnetically almost equivalent 
four Ce$^{+3}$ ions in the primitive cell.  
Namely, the universality class of critical 
fluctuations of such systems belongs to that with the 
dynamical exponent $z$=3 as discussed previously in 
Ref.\cite{hatatani}, so that anomaly of $\rho_{0}$ near QCP is expected 
to become much sharper than that of the conventional AF-QCP with 
the dynamical exponent $z$=2.  This conclusion is consistent with a strong 
pressure dependence of $\rho_{0}$ observed in CeCu$_6$\cite{raymond}, which is 
considered to be located near AF-QCP because CeCu$_{5.9}$Au$_{0.1}$ exhibits 
a non-Fermi liquid behavior characteristic of QCP although its universality 
class has not been identified yet.  
The sharp decrease of $\rho_{0}$ as a function of the pressure, as shown in 
Fig.~2 of Ref.~\cite{raymond}, does not seem to 
be understood as the same mechanism as canonical behaviors of gradual 
decrease of $\rho_{0}$ under pressure which are observed in a series of 
heavy electron systems, such as CeInCu$_2$\cite{kagayama}, 
CeAl$_3$\cite {flouquet}, and so on.  

In the case of QCP associated with a 
valence transition as observed in CeCu$_2$Ge$_2$\cite{onishi}, 
the corresponding Fermi 
liquid effect in (\ref{eq:20}) may give dominant contribution and lead to 
much more pronounced enhancement of $\rho_{0}$ as will be 
discussed elsewhere.  
Such a Fermi liquid correction is related to the higher order perturbations 
with respect to the critical fluctuations.  
Analysis of these higher order terms is left for future study. 
The Fermi liquid correction is also important for 
the enhancement of exchange potential of magnetic impruity near 
the ferromagnetic QCP, and leads to a non-trivial effect~\cite{MMV}.  

The resitivity $\rho_{\rm imp}$ due to impurity scattering is expected to 
have prominent $T$-dependence, in general, arising 
from renormalization of impurity 
potential by the critical fluctuations at around QCP\cite {maezawa}.  
Therefore, one has to be careful when $T$-dependence of the observed 
resistivity is compared to existing theories~\cite{moriya2,rosch}.  

The enhancement of $\rho_{0}$ near the ferromagneitc QCP should work 
to suppress the anisotropic supserconductivity which is expected 
to appear around there~\cite{monthoux}.  One has to remember this effect 
when dicussing the superconductivity induced by critical ferromagnetic  
fluctuations in real metals.  

\section*{Acknowledgements}
One of us (K.M.) acknowledges H. Maebashi for clarifying discussions.  
This work is supported by the Grant-in-Aid for COE Research (10CE2004) 
from Monbu-Kagaku-sho, the Japanese Ministry of Education, Science, Sports, 
Culture and Technology.

\end{document}